\documentstyle[aps,prl,multicol,epsfig]{revtex}
\draft
\begin{document}

\title{Kondo effect induced by a magnetic field} 
\author{M. Pustilnik and L. I. Glazman 
\vspace{1.2mm}}
\address{
Theoretical Physics Institute, University of Minnesota, \\ 
116 Church St. SE, Minneapolis, MN 55455 
\vspace{1mm}}
\maketitle

\begin{abstract}
We study peculiarities of transport through a Coulomb blockade system
tuned to the vicinity of the spin transition in its ground state. Such transitions 
can be induced in practice by application of a magnetic field. Tunneling of 
electrons between the dot and leads mixes the states belonging to the ground 
state manifold of the dot. 
Remarkably, both the orbital and spin degrees of freedom of the electrons
are engaged in the mixing at the singlet-triplet transition point. We present a 
model which provides an adequate theoretical description of recent experiments 
with semiconductor quantum dots and carbon nanotubes.
\end{abstract}


\begin{multicols}{2}
\section{Introduction} 

Quantum dot devices provide a well-controlled object for studying
quantum many-body physics. In many respects, such a device resembles
an atom imbedded into a Fermi sea of itinerant electrons. These
electrons are provided by the leads attached to the dot. The orbital
mixing in the case of quantum dot corresponds to the electron
tunneling through the junctions connecting the dot with leads. Voltage
$V_g$ applied to a gate -- an electrode coupled to the dot
capacitively -- allows one to control the number of electrons $N$ on
the dot. Almost at any gate voltage an electron must have a finite
energy in order to overcome the on-dot Coulomb repulsion and tunnel
into the dot. Therefore, the conductance of the device is suppressed
at low temperatures (Coulomb blockade phenomenon\cite{Leo}). The
exceptions are the points of charge degeneracy. At these points, two
charge states of the dot have the same energy, and an electron can hop
on and off the dot without paying an energy penalty. This results in a
periodic peak structure in the dependence of the conductance $G$ on
$V_g$. Away from the peaks, in the Coulomb blockade valleys, the
charge fluctuations are negligible, and the number of electrons $N$ is
integer.

Every time $N$ is tuned to an odd integer, the dot must carry a
half-integer spin. In the simplest case, the spin is $S=1/2$, and is
due to a single electron residing on the last occupied discrete level
of the dot. Thus, the quantum dot behaves as $S=1/2$ magnetic impurity
imbedded into a tunneling barrier between two massive conductors. It
is known\cite{reviews} since mid-60's that the presence of such
impurities leads to zero-bias anomalies in tunneling
conductance\cite{classics}, which are adequately
explained\cite{Appelbaum} in the context of the Kondo
effect\cite{Kondo}. The advantage of the new experiments\cite{exp1} is
in full control over the ``magnetic impurity'' responsible for the
effect. For example, by varying the gate voltage, $N$ can be
changed. Kondo effect results in the increased low temperature
conductance only in the odd-$N$ valleys. The even-$N$ valleys
nominally correspond to the $S=0$ spin state (non-magnetic impurity),
and the conductance decreases with lowering the temperature. 

However, the enhancement of the low temperature conductance in the even-$N$ 
valleys has been observed recently in both vertical \cite{sasaki} and lateral 
\cite{Weis} GaAs quantum dots, and in transport through carbon nanotubes 
\cite{david}. Unlike the real atoms, the energy separation between the discrete 
states in a quantum dot is fairly small. Therefore, the $S=0$ state of a dot with 
even number of electrons is less robust than the corresponding ground state 
of a real atom. Because of the exchange interactions between electrons residing
on the last doubly occupied level, one of the electrons is promoted to the next 
(empty) orbital state, and thus a triplet spin state of the dot is formed. Apparently,
dots studied in \cite{sasaki} and \cite{Weis} were in a triplet state in the absence 
of magnetic field. On the other hand, experiments with carbon nanotubes \cite{david} 
indicate that in the absence of magnetic field the state with even number of electrons 
is a singlet. 

Application of a magnetic field results in a spin transition in the ground state 
of a quantum dot \cite{sasaki} or carbon nanotube \cite{david}. In the former 
case the Zeeman effect is negligible because of the small $g$-factor in GaAs; 
magnetic field affects mostly the orbital states, making energy of the triplet 
state higher than that of the singlet. In the latter case, the Zeeman contribution 
dominates and can exceed the level spacing, i.e., the magnetic field induces 
the transition to a triplet state. 
Electron tunneling between the dot and leads results in mixing of 
the components of the ground state. The mixing involves spin as 
well as orbital degrees of freedom and yields an enhancement of the low 
temperature conductance through the dot at the transition point. This enhancement 
can be viewed as a {\it magnetic field--induced Kondo effect} \cite{review}.

In Section~\ref{MODEL} a simple model capable of describing the singlet-triplet 
transition in Coulomb blockade systems is introduced.  In Section~\ref{SW} we 
derive an effective low energy Hamiltonian and establish the Fermi-liquid nature of 
its ground state. The Hamiltonian is analyzed by means of renormalization group 
in Sections~\ref{SCALING}-\ref{ZEEMAN}. Generalizations of the model are 
discussed in Appendices~\ref{E&N}-\ref{STAIRCASE}.

Section~\ref{TRANSITION} is relevant for the experiments \cite{sasaki} with 
vertical GaAs quantum dots. In this case the Zeeman energy is very small due to 
an anomalously small electronic $g$-factor in GaAs. The low energy physics in 
this limit is described by a special version of a two-impurity Kondo model \cite{PG}. 
We show that at zero temperature the linear DC conductance is a monotonous 
function of a magnetic field, reaching the unitary limit $\sim 4e^2/h$ at the triplet 
side of the singlet-triplet transition, and decreasing logarithmically at the singlet 
side. The Kondo temperature on the triplet side of the transition falls off rapidly 
from $T_0$ at the transition point down to $T_K^{\rm triplet}$ far away from the 
transition \cite{PG,EN,EN1}. 

The relations between $T_0, T_K^{\rm triplet}$, and the Kondo temperature 
$T_K^{\rm odd}$ in the adjacent Coulomb blockade valleys depend on the 
tunneling amplitudes between the dot and the leads and also on the intradot 
interactions. If the tunneling preserves the orbital symmetry, as it likely occurs 
in vertical quantum dots \cite{sasaki,tarucha}, then 
$T_K^{\rm triplet} \ll T_K^{\rm odd}\lesssim T_0$. Accordingly, at a finite 
temperature $T_K^{\rm triplet} \ll T \lesssim T_0$ the Kondo effect in $N=even$ 
valley of Coulomb blockade can be observed only when the system is tuned to 
the singlet-triplet transition point \cite{sasaki}.  

If the tunneling mixes the orbitals, the difference between $T_K^{\rm triplet}$ 
and $T_K^{\rm odd}$ becomes less pronounced and vanishes if the mixing 
is strong, see Appendix~\ref{MIXING}. The relation between $T_0$ and 
$T_K^{\rm odd}$ is almost independent on the degree of mixing but depends 
strongly on the intradot interactions, see Appendices~\ref{E&N},\ref{MIXING}.

In Section~\ref{ZEEMAN} we discuss the Zeeman splitting--driven transition 
\cite{Zeeman}, relevant for the transport experiments with carbon nanotubes 
\cite{david}. In this system, the $g$-factor is close to its free-electron value, while 
the orbital effects are suppressed due to a small radius of the nanotube. The low 
energy physics in this case is described adequately by the single-channel 
anisotropic Kondo model. Tuning away from the transition points is equivalent 
to applying a finite magnetic field to the conventional Kondo impurity. The DC 
conductance has a peak at the transition point at all values of $T$, with the 
corresponding Kondo temperature being of the same order as $T_K^{\rm odd}$. 

Finally, in Section~\ref{FLIP} we demonstrate that in a generic case the positions 
of the peaks of {\it non-linear} conductance are not symmetric with respect to the 
change of the bias polarity. This effect is weak and can be best seen in double-dot 
systems.

\section{The model}\label{MODEL}

\begin{figure}
\centerline{\epsfxsize=6.5cm
\epsfbox{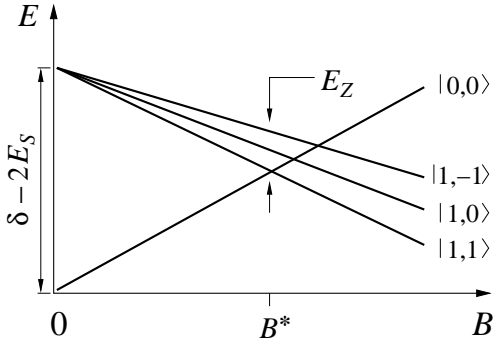}\vspace{1.5mm}}
\caption{Typical picture of the singlet-triplet transition in the ground state of 
a quantum dot.}
\label{crossings}
\end{figure}

We will consider a confined electron system which does not have
special symmetries, and therefore the single particle levels in it are
non-degenerate. In addition, we assume the electron-electron
interaction to be relatively weak (the gas parameter $r_s\lesssim
1$). Therefore, discussing the ground state, we concentrate on the
transitions which involve only the lowest-spin states. In the case of {\it even} 
number of electrons $N$ on the dot, these are states with $S=0$ and $S=1$. 
At a sufficiently large level spacing $\delta\equiv\epsilon_{+1}-\epsilon_{-1}$ 
between the last occupied ($-1$) and the first empty orbital level ($+1$), the 
ground state is a singlet at $B=0$. Finite magnetic field affects the orbital 
energies; if it reduces the difference between the energies of the said orbital 
levels, a transition to a state with $S=1$ may occur, as illustrated at 
Fig.~\ref{crossings}. Such a transition involves rearrangement of two 
electrons between the levels $n=\pm 1$. Out of the six states involved, 
three belong to a triplet $S=1$, and three others are singlets ($S=0$). 
The degeneracy of the triplet states is removed only by Zeeman energy.  
The three singlet states, in general, are not degenerate with each other. To 
describe qualitatively the transition between a singlet and the triplet in the 
ground state, it is sufficient to consider the following Hamiltonian:
\begin{equation}
H_{dot} =\sum_{ns}\epsilon _{n}d_{ns}^{\dagger }d_{ns}
-E_{S}{\bf S}^2 - E_Z S^z 
+E_{C}\left( N-{\cal N}\right) ^{2}.  
\label{Hdot}
\end{equation}
Here, $N=\sum_{s,n}d_{ns}^{\dagger }d_{ns}$ is the total number of
electrons occupying the levels $n=\pm 1$, operator 
\[
{\bf S} = \sum_{nss'}d_{ns}^{\dagger }
\frac {\bbox{\sigma }_{ss'}}{2} d_{ns'}
\] 
is the corresponding total spin of the dot ($\bbox{\sigma}$ are the Pauli matrices), 
and the parameters $E_S$, $E_{C}$, and  
\begin{equation}
E_Z=g_d \mu_B B,
\label{E_Z}
\end{equation}
are the exchange, charging, and Zeeman energies respectively \cite{exchange}.  
We restrict our attention to the very middle of a Coulomb blockade valley 
with an even number of electrons in the dot (that is modelled by setting 
the dimensionless gate voltage ${\cal N}$ to ${\cal N}=2$).  We assume 
that the level spacing $\delta$ is tunable, {\it e.g.}, by means of a magnetic 
field $B$: $\delta=\delta(B)$.  

The lowest energy singlet state and the three components of the competing 
triplet state can be labeled as $\left| S,S^{z}\right\rangle $ in terms of the
total spin $S$ and its $z$-projection $S^z$,
\begin{eqnarray}
&&
|1,1\rangle =d_{+1\uparrow }^{\dagger }d_{-1\uparrow }^{\dagger}|0\rangle, 
\nonumber \\
&&
|1,-1\rangle 
= d_{+1\downarrow}^{\dagger }d_{-1\downarrow }^{\dagger }|0\rangle ,  
\label{Basis} \\
&&
|1,0\rangle 
=\frac{1}{\sqrt{2}}\left( d_{+1\uparrow }^{\dagger}d_{-1\downarrow }^{\dagger }
+ d_{+1\downarrow }^{\dagger }d_{-1\uparrow}^{\dagger }\right) | 0\rangle ,  
\nonumber \\
&&
|0,0\rangle =d_{-1\uparrow }^{\dagger }d_{-1\downarrow}^\dagger |0\rangle,   
\nonumber
\end{eqnarray}
where $|0\rangle $ is the state with the two levels empty.  From (\ref{Hdot}), the 
energies of these states satisfy
\begin{equation}
E_{|S,S^z\rangle} - E_{|0,0\rangle} = K_0 S - E_Z S^z, 
\quad
K_0 = \delta-2E_S
\label{gaps}
\end{equation}
If, for example, $K_0 > 0$ at $B=0$, the ground 
state of the dot in the absence of the magnetic field is a singlet $|0,0\rangle$. Finite 
field shifts the singlet and triplet states due to the orbital effect, and also leads to 
Zeeman splitting of the components of the triplet. As $B$ is varied, the level crossings occur 
(see Fig.~\ref{crossings}).  The first such crossing takes place at $B=B^*$, 
for which
\begin{equation}
\Delta = E_{|0,0\rangle} - E_{|1,1\rangle} = 0.
\label{Bast}
\end{equation}
At this point the two states, $|1,1\rangle$ and $|0,0\rangle$, form a doubly degenerate ground state 
of the dot.

If leads are attached to the dot, the dot-lead tunneling results in the hybridization 
of the degenerate (singlet and triplet) states. The characteristic energy scale $T_0$ 
associated with the hybridization can be in different relations with the Zeeman 
splitting at field $B=B^*$. 

If $E_Z(B^*)\ll T_0$, then the Zeeman splitting can be neglected, and at the 
$B=B^*$ point all {\it four} states (\ref{Basis}) can be considered as degenerate. 
This limit adequately describes a quantum dot formed in a two-dimensional electron 
gas (2DEG) at the GaAs-AlGaAs interface, subject to a magnetic field 
\cite{sasaki,tarucha}. Indeed, the orbital effect of a magnetic field is very 
strong in GaAs due to a smallness of the effective mass, so that a relatively weak 
magnetic field $B^*$ suffices to induce the spin transition. In addition, the 
electronic $g$-factor in GaAs is also small. Both these factors contribute to
the smallness of the Zeeman energy $E_Z (B^*) = g_d \mu_B B^*$ compared to 
$T_0$. This case is studied in details in Section~\ref{TRANSITION}.  

The opposite limit $E_Z(B^*)\gg T_0$ is realized in carbon nanotubes \cite{david}.
In nanotubes, on the one hand, the effect of magnetic field on orbital motion is very 
weak due to their small diameter. On the other hand, the $g$-factor is close to its 
free electron value $g=2$, yielding an appreciable Zeeman splitting even in a magnetic 
field of moderate strength. This limit of the theory is addressed in 
Section~\ref{ZEEMAN}. 

In order to study the transport problem, we need to introduce into the
model the Hamiltonian of the leads and a term that describes the
tunneling. We choose them in the following form: 
\begin{eqnarray}
H_{l}&=&\sum_{\alpha nks}\xi _{ks}c_{\alpha nks}^{\dagger }c_{\alpha nks},
\quad
\xi _{ks} = \xi_k - \frac s2 g_c \mu_B B  
\label{leads} \\
H_{T}&=&\sum_{\alpha nn'ks}t_{\alpha nn'}c_{\alpha nks}^{\dagger
  }d_{n's}+{\rm H.c.}
\label{HT} 
\end{eqnarray}
Here $\alpha =R,L$ for the right/left lead, and $n = \pm 1$ for the two orbitals 
participating in the singlet-triplet transition; $k$ labels states of the continuum 
spectrum in the leads, and $s = \pm 1$ is the spin index. The single-particle energies 
$\xi _{ks}$ include the direct effect of the external magnetic field on the electrons in 
the leads. 
 
In writing (\ref{leads})-(\ref{HT}), we had in mind the vertical dot device 
\cite{sasaki}, where the potential creating lateral confinement of electrons most 
probably does not vary much over the thickness of the dot \cite{tarucha}. Therefore 
we have assumed that the electron orbital motion perpendicular to the axis of the 
device can be characterized by the same quantum number $n$ inside the dot and 
in the leads. 
Presence of two orbital channels $n=\pm 1$ is important for the explanation of 
experiments \cite{sasaki}, in which the orbital effect of the magnetic field 
dominates. 
In the case of large Zeeman splitting, the problem effectively becomes a 
single-channel one \cite{Zeeman}, as we will see in Section~\ref{ZEEMAN}.

\section{Low Energy Hamiltonian}\label{SW}

We will demonstrate the derivation of the effective low energy Hamiltonian under a 
simplifying assumption \cite{PG,EN}
\begin{equation}
t_{\alpha nn'} = t_{\alpha}\delta_{nn'}.
\label{tunneling}
\end{equation}
It has been shown previously \cite{review} that, despite its simplicity, 
Eq.~(\ref{tunneling}) still allows to understand, at least qualitatively, the most 
interesting results of the experiments \cite{sasaki,david}. Some of the features 
not captured by (\ref{tunneling}) will be discussed later on. 

The advantage of (\ref{tunneling}) is that it allows us to perform a rotation in the 
R-L space \cite{AM}
\begin{equation}
\left( 
\begin{array}{c}
\psi _{nks} \\ 
\phi _{nks}
\end{array}
\right) = 
\frac{1}{\sqrt {t_L^2 + t_R^2}} 
\left( 
\begin{array}{cc}
t_{R} & t_{L} \\ 
-t_{L} & t_{R}
\end{array}
\right)
\left( 
\begin{array}{c}
c_{Rnks} \\ 
c_{Lnks}
\end{array}
\right),
\label{rotation}
\end{equation}
after which the $\phi$ field decouples from the dot: 
\begin{equation}
H_T = \sqrt{t_L^2 + t_R^2}\sum_{nks} \psi_{nks}^{\dagger}d_{ns} + {\rm H.c.}
\label{diag}
\end{equation}
The Hamiltonian of the system then acquires a form
\begin{equation}
{\cal H} = H_0(\phi) + H(\psi), 
\label{Hamiltonian}
\end{equation}
where 
$H_0(\phi) = \sum_{nks}\xi_{ks}\phi _{nks}^{\dagger }\phi _{nks}$
with $\xi_{ks}$ given by Eq.~(\ref{leads})
is the free-electron Hamiltonian of the $\phi$-particles, and 
\begin{eqnarray}
&& H(\psi) = H_0 (\psi) + H_{dot} + H_T, 
\nonumber\\
&& H_0(\psi) = \sum_{nks}\xi_{ks}\psi _{nks}^{\dagger }\psi _{nks}
\label{Hpsi}
\end{eqnarray}
describes the $\psi$-particles interacting with the dot via the tunneling 
Hamiltonian (\ref{diag}).

The physical quantity we are interested in is the linear conductance of the system $G$. 
The particle current operator is defined as
\begin{equation}
j = \frac{d}{dt} \frac{1}{2} (N_L-N_R), 
\quad
N_\alpha = \sum_{nks} c_{\alpha nks}^\dagger c_{\alpha nks}
\label{current}
\end{equation}
In order to take a full advantage of the decomposition (\ref{Hamiltonian}), we need to express 
$j$ via $\psi$ and $\phi$ of Eq.~(\ref{rotation}). When the resulting expression is substituted 
into the Kubo formula (see Appendix~\ref{KUBO} for details of the calculation), it yields
\begin{equation}
G = \sum_{ns}g_0 \int d\varepsilon (- df/d\varepsilon) 
\left[- \pi\nu {\rm Im}{\cal T}_{ns}(\varepsilon)\right] .
\label{RESPONSE}
\end{equation}
Here ${\cal T}_{ns}$ are the diagonal elements
of the ${\cal T}$-matrix for the Hamiltonian $H(\psi)$, defined in a usual fashion 
(see Appendix~\ref{KUBO}), 
$f$ is the Fermi function, $\nu$ is the density of states in the leads, and
\begin{equation}
g_0 = \frac {e^2}{2\pi\hbar} \left( \frac{2t_L t_R}{t^2_L + t_R^2} \right)^2 .
\label{g_0}
\end{equation}

In order to bring $H(\psi)$ to the form convenient for further analysis, we explore the 
existence of a one-to-one correspondence between the low energy states of the dot 
(\ref{Basis}) and the states of two fictitious $1/2$-spins \cite{PG}:
\begin{eqnarray}
&& |1,1\rangle  \Longleftrightarrow |\uparrow_1 \uparrow_2 \rangle ,
\nonumber\\
&& |1,-1\rangle \Longleftrightarrow |\downarrow_1 \downarrow_2 \rangle ,
\nonumber\\
&& |1,0\rangle  \Longleftrightarrow \frac{1}{\sqrt{2}}
\left(
|\uparrow_1 \downarrow_2 \rangle + |\downarrow_1 \uparrow_2 \rangle 
\right),
\label{OneToOne}\\
&& |0,0\rangle  \Longleftrightarrow \frac{1}{\sqrt{2}}
\left(
|\uparrow_1 \downarrow_2 \rangle - |\downarrow_1 \uparrow_2 \rangle 
\right),
\nonumber
\end{eqnarray}
This correspondence allows one to represent any operator acting on the states (\ref{Basis}) 
in terms of the spin-$1/2$ operators ${\bf S}_{1,2}$. By comparing directly 
matrix elements, one can check the validity of the following relations: 
\begin{eqnarray}
&&
{\cal P}\sum_{ss'}d_{ns}^{\dagger}\frac{{\bbox \sigma }_{ss'}}{2}d_{n's'}{\cal P} 
\equiv  \frac{\delta_{n,n'}}{2} {\bf S}_{+} 
+ \frac{\delta_{n,-n'}}{2\sqrt 2}({\bf S}_{-} + 2in {\bf T})
\label{Snn} \\
&&
{\cal P}\sum_s d_{ns}^{\dagger }d_{n's}{\cal P} 
\equiv  n\delta_{n,n'} \left[\left({\bf S}_1 \cdot {\bf S}_2 \right) - 1/4 + n \right] 
\label{Cnn}
\end{eqnarray}
Here ${\cal P} = \sum_{S, S^z} |S,S^z\rangle \langle S,S^z|$ 
is a projector onto the low energy multiplet (\ref{Basis}), 
$\bbox{\sigma} = \left(\sigma^x, \sigma^y, \sigma^z \right)$ are the Pauli matrices,
and
\begin{equation}
{\bf S}_{\pm} = {\bf S}_1 \pm {\bf S}_2,
\quad
{\bf T} = {\bf S}_1 \times {\bf S}_2 .
\label{ST}
\end{equation}
Eqs.~(\ref{Snn})-(\ref{Cnn}) should be understood in a sense that the operators in the 
l.h.s. obey the same algebra as those in the r.h.s.  The states (\ref{Basis}) are the 
eigenstates of the operators ${\bf S}_{+}$ and $\left({\bf S}_1 \cdot {\bf S}_2 \right)$, 
while ${\bf S}_{-}$ and ${\bf T}$  describe transitions between the singlet state and the
components of the triplet. Some useful relations involving these operators are listed in 
Appendix \ref{OPERATORS}.

We can now simplify $H(\psi)$ by integrating out the virtual transitions to the states with 
${\cal N}\pm 1$ electrons on the dot. This is done by means of the Schrieffer-Wolff 
transformation or, equivalently, of the second order of the Brilluin-Wigner perturbation 
theory. Using Eqs.~(\ref{Snn})-(\ref{ST}), the result of this procedure is written in terms 
of the operators ${\bf S}_{1,2}$ as\cite{PG}:
\begin{eqnarray}
&&H=H_0(\psi) +K({\bf S}_1\cdot {\bf S}_2) - E_ZS^z_+
+\sum_{n}H_{n}, 
 \label{model}\\
&&H_{n} =J({\bf s}_{nn}\cdot {\bf S}_{+}) 
+ V n\rho_{nn}( {\bf S}_{1}\cdot {\bf S}_{2})   
\nonumber \\
&&\quad\quad\quad\quad
+ \frac{I}{\sqrt{2}}\left[ ({\bf s}_{-n,n}\cdot {\bf S}_{-})
+ 2in({\bf s}_{-n,n}\cdot {\bf T}) \right].   
\label{Hn} 
\end{eqnarray}
Here we introduced the particle and  spin densities in the continuum: 
\begin{equation}
\rho _{nn'}
=\sum_{kk's}\psi _{nks}^{\dagger }\psi _{n'k's},\;
{\bf s}_{nn'}
=\sum_{kk'ss'}\psi_{nks}^{\dagger }
\frac{{\bbox \sigma }_{ss'}}{2}\psi _{n'k's'}.
\label{densities}
\end{equation}
The bare values of the coupling constants are 
\begin{equation}
J=I= 2V= 2(t_L^2 + t_R^2)/E_C. 
\label{bare}
\end{equation}

We did not include into (\ref{model})-(\ref{Hn}) some terms, that are 
irrelevant for the low energy renormalization. The contribution of these terms 
to the conductance is featureless at the energy scale of the order of $T_0$ 
(see the next section), where the Kondo resonance develops. The 
Schrieffer-Wolff transformation also produces a small correction to the 
energy gap $K =  E_{|1,0\rangle}-E_{|0,0\rangle}$ between the states 
$|1,0\rangle$ and $|0,0\rangle$, so that $K \neq K_0$. However, this 
difference is not important, since it only affects the value of the control 
parameter at which the singlet-triplet transition occurs, but not the nature 
of the transition. 

The effective Hamiltonian (\ref{model})-(\ref{Hn}) can be simplified further
by excluding the magnetic field acting on the electrons in the leads. Indeed, 
according to Eq.~(\ref{RESPONSE}), the linear 
conductance is determined by the scattering properties of the $\psi$-particles
in the vicinity of the Fermi level. It is therefore convenient to redefine 
the single particle energies in such a way that being measured from the 
corresponding Fermi levels they are independent of the spin direction 
$\xi_{ks}\to\xi_k$,
\begin{equation}
H_0 = \sum_{nks} \xi_k \psi_{nks}^\dagger \psi_{nks}, 
\label{H_0}
\end{equation}
cf. Eq.~(\ref{Hpsi}). Most of the operators in the 
r.h.s. of (\ref{Hn}) are invariant with respect to the redefinition of energies. 
The only correction that appears is
\begin{equation}
\sum_n J \langle s_{nn}^z \rangle_0 S^z_+ 
= \nu J g_c \mu_B B S^z_+, 
\label{delta_g}
\end{equation}
where $\langle s_{nn}^z \rangle_0$ is ground state average of the spin density 
in the presence of the magnetic field (Note that $J$ here is the {\it bare} value of the 
exchange amplitude). This correction can be readily absorbed into $E_Z$ and
can be cast in the form of a small (since $\nu J \ll 1$) tunneling-induced correction to
the $g$-factor $g_d$ of the dot's electrons [see Eq.~(\ref{E_Z})], indistinguishable from 
other types of corrections \cite{g_factor}. Thus, not quite unexpectedly \cite{TW}, the 
direct Zeeman coupling of the magnetic field to the electrons in the leads has little influence 
on the properties of the system. 

There were several simplifying assumptions involved in the derivation of the effective 
Hamiltonian (\ref{model})-(\ref{Hn}). Nevertheless, we believe it provides an adequate 
description of the low temperature physics in the vicinity of the singlet-triplet transition. 
The dot interacts via tunneling with four species, or channels, of conduction electrons. 
At the same time, the dot behaves like a composite two-spin impurity. Obviously, it takes 
only two channels to screen such an impurity.  By analogy with the multichannel Kondo 
model \cite{NB} one should expect that in the absence of special symmetry no more than 
two channels are coupled to the dot in the effective low energy Hamiltonian. 

In the above derivation of the effective two-channel Hamiltonian we entirely ignored the 
multilevel structure of the dot and also adopted a simplified description 
of intradot Coulomb interaction. The presence of many energy levels in the dot is important 
at the energies above the single particle level spacing $\delta$, while the Kondo physics 
emerges at the energy scale well below $\delta$. These levels result merely in a renormalization 
of the parameters of the effective low energy Hamiltonian \cite{multilevel}. One only needs to 
consider this renormalization for deriving the relation between the parameters 
$t_{\alpha nn'}$ of the low energy Hamiltonian (\ref{Hdot}), (\ref{leads}) and (\ref{HT}) to 
the ``bare'' constants of the model defined in a wide $(\sim\epsilon_F)$ band. On the other hand, 
using the effective low energy Hamiltonian, one can calculate, in principle, the observable 
quantities such as conductance $G(T)$ and other susceptibilities of the system at low 
temperatures ($T\ll\delta$), and establish the relations between them, which is our main 
goal. 

The two-parameter $(E_{C},E_{S})$ description of the intradot Coulomb interaction, see 
Eq.~(\ref{Hdot}), is justified as long as Random Matrix Theory (RMT) is applicable 
\cite{exchange}. Small dots \cite{sasaki,Weis,tarucha}, containing only a few
electrons, can not be described in the framework of RMT. Characterization of the 
interaction in such systems requires introduction of additional parameters. Respectively, the 
part of the effective Hamiltonian which describes the interaction of the dot with itinerant 
electrons contains some additional [compared to Eq.~(\ref{Hn})] constants, 
see Eq.~(\ref{ENmodel}). Similar complications arise if the restriction (\ref{tunneling}) 
on the  tunneling amplitudes is lifted, see Eq.~(\ref{Hn1}). Although the 
Hamiltonians (\ref{Hn1}) and (\ref{ENmodel}) appear to be different and more 
cumbersome than Eq.~(\ref{Hn}), all three models turn out to be identical as far as one is
interested in the low-temperature properties of the system. We will see in the next section that
the three constants, $J$, $V$, and $I$ of the exchange Hamiltonian (\ref{Hn}) grow in the 
course of renormalization. At a sufficiently advanced stage of the renormalization group 
procedure, these constants significantly exceed their starting values. It turns out, 
see Appendices~\ref{E&N} and \ref{MIXING}, that at such a stage the 
Hamiltonians (\ref{Hn1}) and (\ref{ENmodel}) generate the same renormalization 
group (RG) flow as Eq.~(\ref{Hn}) does.

Note that the expression of the current operator (\ref{current}) in terms of the fields 
$\psi$ and $\phi$ depends strongly on the manner in which $\psi,\phi$ are composed from 
the original conduction electrons representing the left and the right leads. Therefore,
lifting the restriction (\ref{tunneling}) results in modification of 
Eqs.~(\ref{RESPONSE}), which express the DC conductance $G$ 
through the properties of $\psi$-particles. In Section~\ref{FLIP} below we study in details 
the corresponding generalization of Eq.~(\ref{RESPONSE}) onto the case of large Zeeman 
energy.

Also note that Eqs.~(\ref{model})-(\ref{Hn}) are valid only in the vicinity of the singlet-triplet 
transition. Far away from the transition, the system's properties can be much more complicated 
compared to the predictions based on (\ref{model})-(\ref{Hn}). An example of such non-universal 
behavior is discussed in Appendix~\ref{STAIRCASE}.

The Hamiltonian (\ref{model})-(\ref{Hn}) is still very complex and its exact treatment 
is very difficult, if possible at all. Fortunately, there exists a strong argument in favor of the 
Fermi-liquid nature of the ground state of (\ref{model})-(\ref{Hn}). Indeed, formally, 
(\ref{model})-(\ref{Hn}) resembles the effective Hamiltonian of the two-impurity Kondo 
model (2IKM), for which $H_n$ is replaced by \cite{ALJ}
\begin{equation}
H_{n}^{2IKM} = J\left({\bf s}_{nn} \cdot {\bf S}_{+}\right) 
+ I \left( {\bf s}_{-n,n} \cdot {\bf S}_{-}\right) 
\label{2IKM}
\end{equation} 
and the parameter $K$ characterizes the strength of the RKKY interaction. 
It is known \cite{ALJ} that 2IKM may undergo a phase transition at some special 
value of $K$. At this point, the system may exhibit non-Fermi liquid properties. However
that this can happen  \cite{ALJ} only if $H$ is invariant with respect to the 
particle-hole transformation
\[
\psi_{nks}\rightarrow s\psi_{n,-k,-s}^\dagger .
\] 
The two extra terms in (\ref{Hn}) as compared to (\ref{2IKM}) violate this invariance.
Therefore, the symmetry that warrants the existence of the non-Fermi liquid state
is absent in our problem.
The logarithmic grow of the two extra terms in (\ref{Hn}) at low energies \cite{PG} 
precludes the possibility that the RG flow passes anywhere near the 2IKM non-Fermi 
liquid fixed point. 

\section{Scaling Analysis}\label{SCALING}

To gain an insight into the low-energy properties of our model, we apply 
the ``poor man's'' scaling technique \cite{PWA}. The procedure consists 
of a successive integration out of the high-energy degrees of freedom and 
yields a set of scaling equations \cite{PG}
\begin{eqnarray}
&& dJ/d{\cal L} = \nu (J^2 + I^2) ,
\nonumber \\  
&& dI/d{\cal L} = 2\nu I( J + V) ,
\label{Scaling} \\  
&& dV/d{\cal L} = 2\nu I^2
\nonumber  
\end{eqnarray}
with the initial conditions (\ref{bare}). Here ${\cal L}=\ln (\delta/D)$, and $\nu$ is 
the density of states in the leads; the initial value of the high energy cutoff $D$ is 
$D=\delta$, see the discussion after Eq.~(\ref{bare}). The scaling procedure also 
generates corrections to $K$. In the following we absorb these corrections in the 
re-defined value of $K$. Equations (\ref{Scaling}) are valid in the perturbative 
regime and as long as
\[
D\gg\left| K\right| ,E_Z,T.
\] 
At certain value of the scale, ${\cal L} = {\cal L}_{0}$, the inverse 
coupling constants simultaneously reach zero:
\[
1/J\left( {\cal L}_{0}\right) 
=1/I\left( {\cal L}_{0}\right) 
=1/V\left({\cal L}_{0}\right) 
=0. 
\]
This can be used to define a characteristic energy scale of the problem $T_0$ through the
equation
\[ 
\ln (\delta/T_{0}) = {\cal L}_{0}.
\]

The value of $T_0$ is obviously non-universal. It is convenient to parametrize $T_0$ by 
a value of constant $c$ in the expression resembling the usual definition of Kondo temperature: 
\begin{equation}
T_0 = \delta\exp [- c/\nu J];
\label{T0}
\end{equation}
here $J = J({\cal L}=0)$ is given by Eq.~(\ref{bare}). For the special choice (\ref{tunneling}) 
of the tunneling amplitudes it was found numerically \cite{PG} that $c = 0.36$. 
It is instructive to compare $T_0$ with the Kondo temperature $T_K^{\rm triplet}$, 
corresponding to $K=-\delta$  at the triplet side of the singlet-triplet transition (see the 
next Section), 
\begin{equation}
T_K^{\rm triplet} = \delta\exp[ -1/\nu J], 
\label{TK_triplet}
\end{equation}
and with the Kondo temperature $T_K^{\rm odd}$ in the adjacent Coulomb blockade 
valleys with $N=odd$. In the latter case, only one of the two orbitals $n=\pm 1$ in the 
dot is involved in the corresponding effective Hamiltonian, which takes the form of 
a single-channel $S=1/2$ Kondo model. The corresponding exchange amplitude is 
$J_{\rm odd} = 4(t_L^2 + t_R^2)/E_C = 2J$, which yields 
\begin{equation}
T_K^{\rm odd} = \delta\exp[ -1/2 \nu J]. 
\label{TK_odd}
\end{equation}
[note that (\ref{TK_triplet}) and (\ref{TK_odd}) are also valid only for the 
model (\ref{tunneling})].
In the limit $\nu J \rightarrow 0$ Eqs.~(\ref{T0})-(\ref{TK_odd}) imply that
\[
T_0 \gg T_K^{\rm odd} \gg T_K^{\rm triplet}.
\]
This inequality explains why the Kondo effect at the singlet-triplet transition point
in the $N=even$ valley appears more pronounced than that in the 
$N=odd$ valleys \cite{sasaki}, and why the experiments with vertical quantum dots
failed to detect the Kondo effect away from the transition when the dot definitely was in 
the triplet state \cite{sasaki}.

The solution of the RG equations (\ref{Scaling}) can be expanded
near ${\cal L} ={\cal L}_{0}$. To the first order in 
${\cal L}_{0}-{\cal L}=\ln D/T_{0}$, we obtain
\begin{equation}
\frac{1}{\nu J} = \frac{\sqrt \lambda}{\nu I}=\frac{\lambda -1}{2\nu V}
=\left( \lambda +1\right) \ln (D/T_{0}),  
\label{ScalingSolution}
\end{equation}
where 
\begin{equation}
\lambda =2+\sqrt{5}\approx 4.2.
\label{lambda}
\end{equation}
It should be emphasized that the constant $\lambda$ is determined by the properties 
of the solutions of (\ref{Scaling}) at large ${\cal L}$ and is universal in the sense that 
it's value does not depend on the bare values of the coupling constants.

The solution (\ref{ScalingSolution}) can be used to calculate the
differential conductance at high temperature $T\gg \left| K\right|,T_{0}$. 
In this regime, the coupling constants are still small, and the conductance can be 
calculated perturbatively from the Hamiltonian (\ref{model})-(\ref{Hn}) with 
renormalized parameters and formula (\ref{RESPONSE}). This results in
\begin{equation}
G/G_0=\frac{A}{\left[\ln (T/T_{0})\right]^2},
\label{transition}
\end{equation}
where $G_0 = 4g_0$ and
\[
A=\left( 3\pi ^{2}/8\right) \left( \lambda +1\right) ^{-2}
\left[1+\lambda +\left( \lambda -1\right) ^{2}/8\right] 
\approx 0.9
\]
is a numerical constant. At low temperature, the RG flow (\ref{Scaling}) terminates 
at either $D\sim|K|$ (which corresponds to the relatively strong orbital effect of the 
magnetic field), or at $D\sim E_Z$ (weak orbital effect). We consider these two cases 
separately.

\section{Singlet-triplet transition \label{TRANSITION}}

In this section, we assume that the Zeeman energy is negligibly small compared 
to all other energy scales. At high temperature $T\gtrsim |K|$, the conductance 
is given by Eq.~(\ref{transition}).  At low temperature $T\lesssim |K|$ and away 
from the singlet-triplet degeneracy point, $|K|\gtrsim T_0$, the RG flow yielding 
Eq.~(\ref{transition}) terminates at energy $D\sim |K|$. On the {\it triplet} side of the
transition, 
\[
K\ll -T_0,
\] 
the two spins ${\bf S} _{1,2}$ are locked together into a
triplet. The system is therefore described by the effective two-channel Kondo model 
with $S=1$ impurity, obtained from Eqs.~(\ref{model})-(\ref{Hn}) by projecting 
out the singlet state and dropping the no longer relevant potential scattering term:
\begin{equation}
H_{\rm triplet} = H_0 + J\sum_{n}({\bf s}_{nn}\cdot {\bf S}).
\label{Htriplet}
\end{equation}
Here $J$ is given by the solution $J({\cal L})$ of Eq.~(\ref{Scaling}), taken at 
${\cal L} = {\cal L}^* = \ln(\delta/|K|)$, which corresponds to $D=|K|$. With further 
decrease of $D$ the renormalization of the exchange amplitude $J$ is governed by 
the standard RG equation \cite{PWA}
\begin{equation}
dJ/d{\cal L} = \nu J^2, 
\quad
{\cal L} > {\cal L}^* .
\label{Tscaling} 
\end{equation}
The solution of Eq.~(\ref{Tscaling}),
\[
1/\nu J({\cal L}) - 1/\nu J({\cal L}^*)  = {\cal L} - {\cal L}^*, 
\]
can be also written as $1/\nu J = \ln (D/T_{K})$ in terms of the running bandwidth $D$ 
and the Kondo temperature 
\begin{equation}
T_K(K) = |K|\exp\left[ -1/\nu J({\cal L}^*) \right]. 
\label{Tk_triplet}
\end{equation}
Substituting here the asymptote of $J({\cal L}^*)$ from 
Eq.~(\ref{ScalingSolution}) we obtain  
\begin{equation}
\frac{T_K}{T_0} = \left(\frac{T_0}{|K|}\right)^\lambda . 
\label{TKondo_triplet}
\end{equation}
This result is universal and is valid in the vicinity of the transition point. To clarify 
what does the "vicinity`` mean, we again rely on the model (\ref{tunneling}),
for which Eqs.~(\ref{Scaling}) are valid for all $|K| \lesssim \delta$. Expansion of 
the solution of Eqs.~(\ref{Scaling}) near the weak coupling fixed point ${\cal L}=0$
and the substitution of $J({\cal L}^*)$ into (\ref{Tk_triplet}) results in the 
large-$|K|$ asymptote of $T_K(K)$: 
\begin{equation}
\frac{T_K(K)}{T_K^{\rm triplet}} = \frac{\delta}{|K|},
\label{TKK_triplet}
\end{equation}
where $T_K^{\rm triplet}$ is given by Eq.~(\ref{TK_triplet}). Comparison of 
(\ref{TKondo_triplet}) and (\ref{TKK_triplet}) shows that the two asymptotes 
match at
\[
|K|/T_0 = (\delta/T_0)^\mu, 
\quad 
\mu= \frac{1/c- 2}{\lambda - 1} \approx 0.24,
\]
which gives the upper limit of the validity of Eq.~(\ref{TKondo_triplet}): 
$|K|/T_0 \ll (\delta/T_0)^\mu$. 
From Eqs.~(\ref{TKondo_triplet}),(\ref{TKK_triplet}) follows that the larger is 
$|K|$ the smaller $T_K$ is.  In fact, the universal part $T_K(K)$, Eq.~(\ref{TK_triplet}),  
suffices to describe a fall of $T_K$ by an order of magnitude for realistic values 
of the parameters for a vertical quantum dot device \cite{review}.

For a given $|K|\gtrsim T,T_0$ the linear conductance can be cast into the scaling 
form,
\begin{equation}
G/G_{0}=F\left( {T}/{T_{K}}\right) 
\label{triplet}
\end{equation}
where $F\left( {x}\right) $ is a smooth function that interpolates
between $F\left( x\gg 1\right) =\left( \pi ^{2}/2\right) (\ln x) ^{-2}$
and $F\left( 0\right) =1$. It coincides with the scaled resistivity
\[
F(T/T_K)=\rho (T/T_K)/\rho(0)
\] 
for the symmetric two channel $S=1$ Kondo model.  The conductance at
$T=0$ (the unitary limit value) is $G_{0} = 4g_0$, see Eq.~(\ref{g_0}). 
Equation (\ref{triplet}) remains valid not too far away from the transition point, 
where the universal description (\ref{model})-(\ref{Hn}) is applicable. At large 
$|K|$ the conductance exhibit additional features (an example is discussed in 
Appendix \ref{STAIRCASE}). However, smallness of $T_K$ renders 
these features difficult to observe.
 
\begin{figure}[tbp]
\centerline{\epsfxsize=6.5cm
\epsfbox{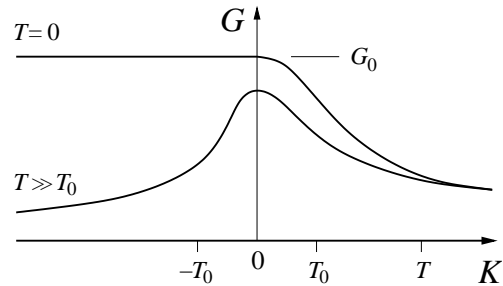}\vspace{1.5mm}}  
\caption{
Linear conductance near a singlet-triplet transition.  At high
temperature $G$ exhibits a peak near the transition point. At low
temperature $G$ reaches the unitary limit at the triplet side of the
transition, and decreases monotonously at the singlet side. The two
asymptotes merge at $K\gg T,T_0$.}
\label{overall}
\end{figure}

On the {\it singlet} side of the transition, 
\[
K\gg T_0,
\] 
the scaling
terminates at $D\sim K$, and the low-energy effective Hamiltonian is
\[
H_{\rm singlet} = H_0 - \frac{3}{4}V\sum_{n}n\rho _{nn}, 
\]
where $V$ is the renormalized scattering amplitude $V({\cal L})$ taken
at ${\cal L}={\cal L}^* = \ln(\delta/|K|)$. The temperature dependence of the
conductance saturates at $T\lesssim K$, reaching the value
\begin{equation}
G/G_0=\frac {B}{\left[\ln (K/T_0)\right] ^2},
\quad 
B=\left( \frac{3\pi }{8}\frac{\lambda -1}{\lambda +1}\right)^2
\approx 0.5. 
\label{singlet}
\end{equation}
This result is universal in the vicinity of the transition.

At $T=0$ Eqs.~(\ref{triplet}) and (\ref{singlet}) predict different dependence 
on the parameter $K$ which is used for tuning through the transition. At positive 
$K$, conductance decreases with the increase of $K$, while at $K\ll -T_0$ 
conductance $G$ is independent of $K$, see Fig.~\ref{overall}.  
While the conductance $G$ is a monotonous function of $K$ at $T = 0$, it develops 
a {\it peak} at the singlet-triplet transition point $K= 0$ at high temperature $T \gg T_0$. 

\section{Transition Driven by Zeeman Splitting \label{ZEEMAN}}

\begin{figure}
\centerline{\epsfxsize=5cm
\epsfbox{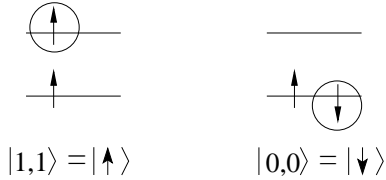}\vspace{2mm}}
\caption{ 
The ground state doublet in case of a large Zeeman splitting. The
states $|1,1\rangle$ and $|0,0\rangle$ differ by flipping a spin of a
single electron (marked by circles).}
\label{states}
\end{figure}

If the Zeeman energy is large, the RG flow (\ref{Scaling}) terminates
at $D\sim E_Z$. The effective Hamiltonian, valid at the energies
$D\lesssim E_Z$ is obtained by projecting (\ref{model})-(\ref{Hn})
onto the states $|1,1\rangle$ and $|0,0\rangle$. These states differ
by a flip of a spin of a single electron (see Fig.~\ref{states}), and
are the counterparts of the spin-up and spin-down states of $S=1/2$
impurity in the conventional Kondo problem. It is therefore 
convenient to switch to the notations \cite{Zeeman}
\begin{equation}
|1,1\rangle = | \uparrow \rangle, 
\quad
|0,0\rangle = | \downarrow \rangle, 
\label{spins}
\end{equation}
and to describe the transitions between the two states in terms of the
(pseudo)spin operator 
\[
\widetilde {\bf S} = \frac{1}{2} \sum_{ss'} 
|s\rangle \bbox{\sigma}_{ss'}\langle s'| ,
\]
built from the states (\ref{spins}).

Among the various operators in (\ref{model})-(\ref{Hn}) that act on
the states of the dot only the following have non-zero matrix elements between the 
states (\ref{spins}): 
\begin{eqnarray*}
&& {\cal P}'S_+^z {\cal P}' = \widetilde{S}^z +1/2, 
\quad
{\cal P}' ({\bf S}_1 \cdot {\bf S}_2) {\cal P}' = \widetilde{S}^z -1/4, 
\\
&& {\cal P}'S_-^\pm {\cal P}' = -{\sqrt 2}\widetilde{S}^\pm, 
\quad
{\cal P}' T^\pm {\cal P}' = \pm \frac{i}{\sqrt 2}\widetilde{S}^\pm, 
\end{eqnarray*}
where ${\cal P}' = \sum_{s}|s\rangle \langle s|$. 
Using these relations, we obtain from (\ref{model})-(\ref{Hn})
\begin{eqnarray}
H=&&\sum_{nks}\xi _{k}\psi _{nks}^{\dagger }\psi _{nks}
- \Delta \widetilde{S}^z 
\nonumber\\
&& +\sum_{n} \left [ 
Js^z_{nn} \left(\widetilde{S}^z +1/2\right)
+ Vn\rho_{nn} \left(\widetilde{S}^z -1/4\right)
\right ]
\label{projected} \\
&& -I\left( s^+_{1,-1}\widetilde{S}^-   +  {\rm H.c.}\right),
\nonumber
\end{eqnarray}
where $\Delta = E_Z - K$, see Eq.~(\ref{Bast}). It is now convenient to 
transform (\ref{projected}) to a form which is diagonal in the orbital 
indexes $n$. This is achieved by relabeling the fields according to
\begin{eqnarray}
&&\psi_{+1,k,\uparrow}= a_{k,\uparrow},
\;
\psi_{-1,k, \downarrow}= -a_{k,\downarrow}, 
\nonumber \\
&&\psi_{-1, k,\uparrow}= b_{k,\uparrow},
\;
\psi_{+1,k, \downarrow}= -b_{k, \downarrow},
\label{fields}
\end{eqnarray}
which yields
\begin{eqnarray}
H=&& H_0 - \Delta \widetilde{S}^z 
\nonumber \\
&&+ V_a s^z_a
 + J_z s^z_a \widetilde{S}^z  
+ \frac{1}{2} J_\perp \left ( s^+_a \widetilde{S}^- + s^-_a \widetilde{S}^+\right ) 
\label{kkondo} \\
&& +  V_bs^z_b +   J'_z s^z_b \widetilde{S}^z ,
\nonumber
\end{eqnarray}
where $H_0$ is a free-particle Hamiltonian for $a,b$ electrons, and  
${\bf s}_a=\sum_{kk'ss'}a_{ks}^{\dagger }
\left(\bbox{\sigma }_{ss'}/2\right)a _{k's'}$ 
is the spin density for $a$-electrons (with a similar definition for ${\bf s}_b$).  The 
coupling constants in (\ref{kkondo}),
\begin{eqnarray}
&& V_a = (J-V)/2,\; J_z= J+2V, \; J_\perp = 2I, 
\nonumber \\
&& V_b = (J+V)/2, \; J'_z= J-2V
\label{amplitudes} 
\end{eqnarray}
are expressed through the solutions of the RG equations
(\ref{Scaling}) taken at ${\cal L} =  {\cal L}^{*} = \ln(\delta/E_Z)$. 

Eq.~(\ref{kondo}) contains unusual terms $V_{i} s^z_{i}$, which have a
meaning of a magnetic field acting {\it locally} on the conduction electrons at
the impurity site.  The appearance of these terms is a price of the finiteness of the 
magnetic field, which breaks spin-rotational invariance. Fortunately, these terms
do not change significantly properties of the model. Their main effect is to produce a
small correction to $\Delta$, through creating a non-zero expectation value
$\langle s^z_{i} \rangle$.  This correction is not important, since it merely shifts the 
degeneracy point. In addition, $V_{i} s^z_{i}$ lead to insignificant corrections
to the densities of states \cite{Zeeman}. In the following, we will discard these terms.

The scaling equations for the coupling constants in (\ref{kkondo}) take the standard
form \cite{PWA}
\begin{equation}
dJ_z/d{\cal L} = \nu J_\perp^2,
\;
dJ_\perp/d{\cal L} = \nu J_z J_\perp, 
\;
{\cal L} > {\cal L}^{*}; 
\label{RGzeeman} 
\end{equation}
$J'_z$ is not renormalized in this order of perturbation theory, but acquires a negative 
correction in the next order:
$d J'_z/d{\cal L} = - \frac{1}{2} \nu^2 J_\perp^2 J'_z$. In addition, $J_z'$ initially 
is smaller than $J_z$ (for $E_Z = \delta$, for example, $J_z'$  simply vanishes).  
Therefore the $b$-dependent part of (\ref{kkondo}) can be ignored, and the 
resulting effective Hamiltonian is that of a single-channel $S=1/2$ anisotropic Kondo 
model \cite{Zeeman}.

The Kondo temperature $T_K$ depends on $E_Z$. For the model (\ref{tunneling}), as 
follows from Eqs.~(\ref{bare}) and (\ref{amplitudes}), the exchange at $E_Z = \delta$ is 
isotropic:  $J_z = J_\perp=2J$ (where $J$ is given by (\ref{bare})). Therefore the 
Kondo temperature $T_K^{\rm Zeeman}$ corresponding to $E_Z = \delta$ satisfies
\[
T_K^{\rm Zeeman} = T_K^{\rm odd},
\]
see Eq.~(\ref{TK_odd}). At lower $E_Z$ the Kondo temperature becomes larger. 
Interestingly, the isotropy of the exchange in (\ref{kkondo}) is preserved in the regime 
when Eqs.~(\ref{Scaling}) can be linearized near ${\cal L} = 0$. The corresponding 
asymptote of $T_K(E_Z)$ is
\begin{equation}
\frac{T_K(E_Z)}{T_K^{\rm Zeeman}} = \left(\frac{\delta}{E_Z}\right)^{1/2}
\label{weak}
\end{equation}
When $E_Z$ approaches $T_0$ one can use Eqs.~(\ref{ScalingSolution}) to calculate 
initial conditions for (\ref{RGzeeman}). From (\ref{ScalingSolution}) and 
(\ref{amplitudes}) follows that the difference between $J_z$ and $J_\perp$ in this regime is 
very small: $J_z/J_\perp = {\sqrt \lambda}/2 \approx 1.02$. Therefore, the anisotropy of the 
exchange can be neglected in the universal regime as well, and we obtain
\begin{equation}
\frac{T_K}{T_0} = \left(\frac{T_0}{E_Z}\right)^{1/\lambda}.
\label{strong}
\end{equation}
The asymptotes (\ref{weak}) and (\ref{strong}) match at
\[
E_Z/T_0 = (\delta/T_0)^{\mu'},
\quad
\mu' = \frac{3-1/c}{1-2/\lambda} \approx 0.42,
\] 
which separates their regions of applicability. The dependence of $T_K$ on $E_Z$ appears to 
be much weaker than the dependence of $T_K$ on $K$ in the previous section, and practically 
saturates in the regime (\ref{strong}).
 
The linear conductance is given by [cf. Eq.~(\ref{RESPONSE})] 
\begin{equation}
G = \sum_{s}g_0 \int d\varepsilon (- df/d\varepsilon) 
\left[- \pi\nu {\rm Im}{\cal T}_{s}^a (\varepsilon)\right],
\label{landauer}
\end{equation}
where ${\cal T}_{s}^a$ is the $\cal T$-matrix for the $a$-particles 
with spin $s$; ${\cal T}_{s}^a$ does not depend on $s$ if the term $V_as_a^z$ in (\ref{kkondo}) 
is neglected. 
The distance to the degeneracy point $\Delta$ in Eq.~(\ref{kkondo}) plays the part of the Zeeman 
splitting of the impurity spin in the conventional Kondo effect. At $\Delta =0$ the Kondo resonance 
develops: Regardless the initial anisotropy of the exchange constants in Eq.~(\ref{kondo}), the 
conductance in the universal regime (when $T$ approaches the Kondo temperature $T_K$ or lower) 
is given by
\begin{equation}
G=G_0 f(T/T_K),
\label{Gzeeman}
\end{equation}
where $f(x)$ is a smooth function interpolating between $f(0)=1$ and 
$f(x\gg 1) = (3\pi^2/16)(\ln x)^{-2}$. Function $f(T/T_K)$ coincides with the scaled resistivity 
for the single channel $S=1/2$ Kondo model and its detailed shape is known \cite{CHZ}. Note 
that in (\ref{Gzeeman}) the zero-temperature conductance $G_0 = 2g_0$ is by a factor of 2 smaller 
than that in Eqs.~(\ref{transition}),(\ref{triplet}),(\ref{singlet}).

At finite $\Delta \gg T_K$, the scaling trajectory (\ref{RGzeeman}) terminates
at $D \sim \Delta$. Below this energy scale the spin-flip scattering processes are suppressed. 
As a result, at $T \ll \Delta$ the linear conductance 
is given by
\begin{equation}
G = G_0 \frac{\pi^2/16}{\left[\ln(\Delta/T_K)\right]^2}
\label{largeDelta}
\end{equation}
and is independent of temperature.
\section{Spin filtering and asymmetry of the non-linear conductance caused 
by Zeeman splitting}
\label{FLIP}

The low-temperature properties of the system exhibiting the Zeeman--splitting--driven Kondo 
effect can be described by a single-channel anisotropic Kondo Hamiltonian 
\begin{eqnarray}
{\cal H} = && H_0(\phi) + H_0(\psi) - \Delta \widetilde{S}^z 
\nonumber\\
&& +  Us^z + J_z s^z  \widetilde{S}^z 
+ \frac{J_\perp}{2} \left(s^+ \widetilde{S}^- + s^- \widetilde{S}^+ \right) ,
\label{kondo} 
\end{eqnarray}
where 
${\bf s} = \sum_{kk'ss'}\psi_{ks}^\dagger \left({\bbox \sigma}_{ss'}/2\right)\psi_{k's'}$. 
This description is valid regardless the assumption (\ref{tunneling}). 
The operators $\psi_{ks}$ are certain linear combinations of the conduction electron 
operators from the left and right leads. When the restriction (\ref{tunneling}) is lifted,
the coefficients of such linear relation are $s$-dependent \cite{Zeeman}:
\begin{equation}
\left(
\begin{array}{c}
\psi _{ks} \\
\phi _{ks}
\end{array}
\right) 
= 
\left(
\begin{array}{cc}
\cos \vartheta_s & \sin \vartheta_s 
\\
- \sin \vartheta_s & \cos \vartheta_s 
\end{array}
\right) \left(
\begin{array}{c}
c_{Rks} \\
c_{Lks}
\end{array}
\right). 
\label{LRrotation}
\end{equation}
Perhaps, the simplest possible generalization of (\ref{tunneling}) yielding
$\vartheta_{+1} \neq \vartheta_{-1}$ 
in (\ref{LRrotation}) is 
$t_{\alpha n n'} = t_{\alpha n} \delta_{nn'}$ 
[see Appendix~\ref{STAIRCASE}, Eq.~(\ref{tunnelingNEW})], in which case 
$\tan\vartheta_s = t_{Ls}/t_{Rs}$. 

The $s$-dependence of the coefficients in (\ref{LRrotation}) may lead to interesting effects.
We consider the linear regime first. It is convenient to split the current operator 
(\ref{current}) into two parts,
\[
j = j_0 + \delta j ,
\]
which are respectively bi-linear and quadratic in the operators $\psi$ and $\phi$.
Using Eq.~(\ref{LRrotation}) we obtain
\begin{eqnarray}
&& 
j_0  = \frac{d}{dt} \frac {1}{2}\sum_{ks} \sin(2\vartheta_s) 
\left(\psi_{ks}^\dagger \phi_{ks} + \phi_{ks}^\dagger \psi_{ks}\right)
\label{j_0} \\
&&
\delta j =  \frac {d}{dt} \frac {\eta}{2} \sum_s s\psi_{ks}^\dagger \psi_{ks},
\quad
\eta = \sum_s s \sin^2 \vartheta_s .
\label{eta}
\end{eqnarray}

The operator $\delta j$ makes no contribution to the DC conductance. Indeed,
since $\frac {1}{2} \sum_s s\psi_{ks}^\dagger \psi_{ks} + \widetilde{S}^z$
commutes with ${\cal H}$, Eq.~(\ref{eta}) can be re-written as
\begin{equation}
\delta j = - \eta \frac{d}{dt} \widetilde{S}^z.
\label{delta_j}
\end{equation}
Being averaged over time, $\delta j$ yields $0$: 
\begin{eqnarray*}
\overline{\langle\delta j\rangle} 
&& = \lim_{t_0\rightarrow \infty}\frac{1}{2t_0}
\int_{-t_0}^{t_0} dt \langle \delta j (t)\rangle 
\\
&& = 
\lim_{t_0\rightarrow \infty}\frac {\eta}{2t_0} 
\left\langle \widetilde{S}^z(-t_0) - \widetilde{S}^z(t_0)\right\rangle
= 0 
\end{eqnarray*}
because operator $\widetilde{S}^z$ is bounded, 
$-1/2\leq \langle\widetilde{S}^z (t)\rangle \leq 1/2$. Thus, the contribution 
$\delta G$ from $\delta j$ to the DC conductance {\it vanishes identically}. 

The contribution from $j_0$ is evaluated in the same way as in 
Appendix~\ref{KUBO} with the result
\begin{equation}
G = \sum_{s}g_0^s \int d\varepsilon (- df/d\varepsilon)  
\left[- \pi\nu {\rm Im}{\cal T}_{s} (\varepsilon)\right],
\label{landauerNEW} 
\end{equation}
where
\[
\quad g_0^s = \frac{e^2}{2\pi \hbar}\sin^2 (2\vartheta_s) 
\]
and ${\cal T}_{s}$ is ${\cal T}$-matrix for $\psi$-particles with spin $s$.
Equation (\ref{landauer}) is identical to Eq.~(\ref{landauerNEW}), except for 
$s$-dependence of $g_0^s$. Therefore, the conductance at sufficiently low 
temperature is again given by Eqs.~(\ref{Gzeeman}) and (\ref{largeDelta}) with 
$G_0=\sum_s g_0^s$. 

At $T=0$ the electrons scatter from the dot elastically \cite{N}, that is, without 
flip of their spin. In this regime the inequality $g_0^{+1} \neq g_0^{-1}$ means 
that the probability of transmission through the dot depends on the direction of 
the spin of an incoming electron. In other words, the system acts as a 
{\it spin filter} \cite{Zeeman}.

\begin{figure}[tbp]
\centerline{\epsfxsize=5cm
\epsfbox{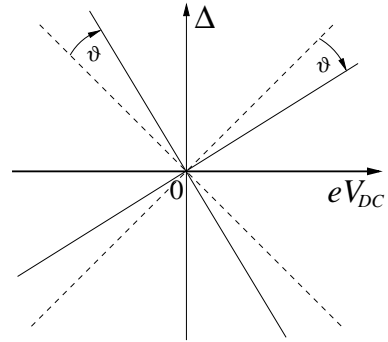}\vspace{1.5mm}}
\caption{Position of the peaks of differential conductance at $(eV_{DC},\Delta)$-plane. The dashed
lines correspond to $eV_{DC} = \pm \Delta$. The angle 
$\vartheta \approx \chi/2 =  \eta\nu J_z/4 \ll 1$ 
can be either positive or negative, depending on the sign of the non-universal parameter $\eta$. }
\label{PEAKS}
\end{figure}

The main effect for the {\it non-linear} differential conductance $dI/dV_{DC}$ 
consists in the violation of symmetry with respect to the bias sign. To see this, it is 
sufficient to calculate the conductance in the leading (second) order of the perturbation 
theory in $J_\perp$. In this order, tunneling of an electron through the dot is 
accompanied by the transition of the dot between the states $|\uparrow\rangle$ and 
$|\downarrow\rangle$, see Eq.~(\ref{spins}). The difference of the corresponding 
energies, $E_{|\uparrow\rangle} - E_{|\downarrow\rangle}$, depends on the average 
value of the operator $s^z $ through the term $J_z s^z  \widetilde{S}^z$ in the 
Hamiltonian (\ref{kondo}). This average is not zero due to a {\it finite} DC voltage 
bias $V_{DC} $, which is described by an addition of a term  
\begin{equation}
H_V = \frac{eV_{DC}}{2} (N_L - N_R)
\label{bias}
\end{equation}
to the Hamiltonian (\ref{kondo}). Using (\ref{LRrotation}) and (\ref{bias}) we obtain 
\begin{eqnarray*}
\langle s^z \rangle && 
=  \sum_{ks}\frac{s}{2} \left[
\langle c_{Rks}^\dagger c_{Rks}\rangle \cos^2 \vartheta_s
+
\langle c_{Lks}^\dagger c_{Lks}\rangle \sin^2 \vartheta_s 
\right]   
\\ &&
= - \frac\eta2 \nu eV_{DC} ,
\end{eqnarray*}
where $\eta$ is given by (\ref{eta}). Therefore, the energy difference between the 
states (\ref{spins}) acquires a bias-dependent correction: 
\begin{eqnarray}
&&
E_{|\downarrow\rangle} - E_{|\uparrow\rangle} = \Delta +  \delta\Delta,
\label{deltaDelta}\\
&&
\delta\Delta = J_z \langle s^z \rangle = \chi eV_{DC}  ,
\quad
\chi = \eta \nu J_z/2.
\nonumber
\end{eqnarray}
This correction is similar to that caused by the magnetic field applied to the leads, see 
Eq.~(\ref{delta_g}). But there is an important difference: $\delta\Delta$ in Eq.~(\ref{deltaDelta})
is due to {\it non-equilibrium}.  
Obviously, the contribution to the non-linear conductance we discuss is suppressed if 
$|eV_{DC} | < E_{|\uparrow\rangle} - E_{|\downarrow\rangle}$. Using Eq.~(\ref{deltaDelta}) we 
find for the interval of suppression
\begin{equation}
-\frac{\Delta}{1+\chi} <eV_{DC} < \frac{\Delta}{1-\chi}.
\label{supression}
\end{equation}
Similar to the conventional Kondo problem \cite{Appelbaum} 
higher-order calculation yields the differential conductance which peaks at the ends of the 
interval (\ref{supression}). Therefore the position of peaks is not symmetric with respect to 
the change of the bias polarity $V_{DC} \rightarrow -V_{DC} $, see Fig.~\ref{PEAKS}.

The feasibility of an experimental observation of this effect obviously depends on the 
value of the parameter $\eta$.  The largest possible value $|\eta|=1$ is reached in a strongly 
asymmetric setup. An example of such setup consists of level $n=+1$, see Fig.~\ref{states}, 
coupled only to the left lead, while the level $n=-1$ is coupled only to the right lead. Such a 
control over the tunneling amplitudes to the individual energy levels in the dot is very difficult to 
achieve with a single-dot device. But one can instead think of a system of two very small quantum 
dots connected {\it in series} \cite{loss}. If the gate voltages are tuned properly, the double-dot 
system is equivalent to a single dot with $N=even$. In particular, the situation is possible when 
the wave functions of the upper and lower levels in Fig.~\ref{states} are localized mostly in the 
left and in the right dots respectively. The degree of the ``localization" and, therefore, the value of 
the parameter $\eta$ is controlled by the strength of the tunneling coupling between the dots. 

Note that similar non-linear effects occur also in the case of a small Zeeman splitting, $E_Z \ll T_0$, 
considered in Section~\ref{TRANSITION}. At a finite temperature and when the system is tuned 
to the singlet-triplet transition point, the differential conductance exhibits a peak at zero bias.  
The peak is split in two when the magnetic field is tuned away from the transition. The location 
of the split peaks is given by the energy balance condition for activation of the interorbital 
scattering processes: $\left|eV_{DC}\right| = K + \delta K$. Here 
$\delta K \propto eV_{DC}$ is a non-universal correction analogous to $\delta\Delta$ in 
(\ref{deltaDelta}), which appears due to a non-zero average of the operator $\sum_n n\rho_{nn}$ 
in Eq.~(\ref{Hn}) in the presence of a finite bias. 

\section*{Conclusion}
In this paper we studied the transport through a quantum dot tuned to the
singlet-triplet transition point. Such a transition in the ground state of an 
isolated dot can be reached by applying a magnetic field to it. The low 
temperature properties of a quantum dot attached to conducting leads can be 
described by a version of a two-impurity Kondo model. Depending 
on the value of Zeeman energy, two distinct physical situations are realized. 

If Zeeman energy is small, the zero-temperature linear conductance reaches the 
quantum limit at the triplet side of the transition. At finite temperature, however, 
the conductance reaches a maximum at the transition point. This happens due 
to the extra degeneracy of the ground state multiplet of the dot at this point, 
which causes an increase of the characteristic temperature below which the 
conductance is approaching the unitary limit.

If Zeeman energy is large,  the ground state of the dot is degenerate only at the 
transition point. In this limit the model can be further reduced to the single-channel 
anisotropic Kondo Hamiltonian. The conductance exhibits a peak at the point of 
the transition at arbitrary low temperature.

The two limits of the theory are in a good agreement with the results of recent 
experiments with vertical quantum dots and carbon nanotubes.

\section*{Acknowledgments}
This work was supported by NSF under Grants DMR-9812340 and DMR-9731756. 
We benefited from discussions with 
D. Cobden, M. Eto, K. Kikoin, L. Kouwenhoven, A. Ludwig, 
A. Luther, K. Matveev, Yu. Nazarov, S. Tarucha, A. Tsvelik, 
M. Voloshin, J. Weis, and A. Zvyagin.

\appendix
\section{Some properties of the operators ${\bf S}_\pm$ and ${\bf T}$}
\label{OPERATORS}

Using 
$\left[ S_i^\alpha , S_j^\beta \right] 
= \delta_{ij}\sum_\gamma \epsilon_{\alpha \beta \gamma} S_i^\gamma $,
it is straightforward to show that the operators 
${\bf S}_\pm = {\bf S}_1 \pm {\bf S}_2$ and 
${\bf T} = {\bf S}_{1}\times {\bf S}_{2}$ obey the commutation relations 
\begin{eqnarray*}
\left({\bf S}_{-}\cdot {\bf T}\right) - \left({\bf T}\cdot {\bf S}_{-}\right) 
&&
= 4i \left({\bf S}_{1}\cdot {\bf S}_{2}\right), 
\\
\left[ {\bf S}_{-},({\bf S}_1\cdot {\bf S}_2) \right] &&= -2i{\bf T}, 
\\
\left[ {\bf T},({\bf S}_1\cdot {\bf S}_2) \right] &&=\frac{i}{2}{\bf S}_{-} .
\end{eqnarray*}
Using these relations one can check that ${\bf S}_{-}$ and ${\bf T}$ are related by 
means of the unitary transformation
\[
2{\bf T} = 
e^{-i\frac{\pi}{2}({\bf S}_1\cdot {\bf S}_2)}
{\bf S}_{-} 
e^{i\frac{\pi}{2}({\bf S}_1\cdot {\bf S}_2)} .
\]
Since $\left({\bf S}_{1}\cdot {\bf S}_{2}\right)$ commutes with ${\bf S}_{+}$, 
application of this transformation to the identity
\[
\left[S_{\pm}^{\alpha},S_{-}^{\beta}\right] 
=\sum_{\gamma}i\epsilon _{\alpha \beta \gamma }S_{\mp}^{\gamma},
\]
gives at once 
\[
\left[S_{+}^{\alpha},T^{\beta}\right] 
=\sum_{\gamma}i\epsilon _{\alpha \beta \gamma}T^{\gamma },
\quad
\left[T^{\alpha},T^{\beta}\right] 
=\frac{1}{4}\sum_{\gamma} i\epsilon _{\alpha \beta \gamma}S_{+}^{k} . 
\]

\section{Derivation of Eq.~(\ref{RESPONSE})}
\label{KUBO}

Using Eq.~(\ref{rotation}), the particle current operator $j$ is written as
\[
j = \frac {t_L t_R}{t^2_L + t_R^2} \sum_{nks} \frac{d}{dt} A_{nks} + {\rm H.c.},
\quad
A_{nks} = \psi_{nks}^\dagger \phi_{nks} 
\] 
Substitution into the Kubo formula 
\begin{equation}
G = \lim_{\omega \rightarrow 0} \frac{e^2}{\hbar}\frac{1}{\omega } \int_0^\infty
dt e^{i\omega t}
\left\langle \left [ j(t), j(0)\right ]\right\rangle ,
\label{kubo}
\end{equation}
and integration by parts yield
\begin{equation}
G = - \lim_{\omega \rightarrow 0}g_0 \pi \omega \sum_{nks} {\rm Im} \Pi_{nks}(\omega),
\label{kubo1}
\end{equation}
where $g_0$ is given by Eq.~(\ref{g_0}) and 
$\Pi_{nks}(\omega) = \int dt e^{i\omega t} \Pi_{nks}(t)$ is the Fourier-transform of the 
retarded correlation function
\[
\Pi_{nks}(t) = -i\theta(t)\left\langle\left[A_{nks}(t), A_{nks}^\dagger(0)\right]\right\rangle
\]
One can express $\Pi_{nks}(\omega)$ via the retarded Green functions of
the $\psi$ and $\phi$ particles as
\begin{eqnarray}
{\rm Im} \Pi_{nks}(\omega) 
= \frac{1}{\pi} \int d\epsilon && \left[f(\epsilon + \omega) - f(\epsilon)\right] 
\label{Pi}\\
&&
\times\, {\rm Im} {\cal G}^{0}_{nks}(\epsilon) {\rm Im} {\cal G}_{nks}(\epsilon +\omega),
\nonumber 
\end{eqnarray}
where $f(\epsilon)$ is the Fermi function, and ${\cal G}_{nks}(\varepsilon)$ is the 
Fourier-transform of 
\[
{\cal G}_{nks}(t) = -i\theta(t)\langle\{ \psi_{nks}(t),\psi_{nks}^\dagger(0)\}\rangle .
\]
In writing of Eq.~(\ref{Pi}) we took into account that the Green function of the $\phi$-particles 
coincides with the unperturbed (without interactions) value of ${\cal G}_{nks}$,
\[
{\cal G}_{nks}^0(\varepsilon) = (\varepsilon - \xi_{ks} + i0)^{-1}.
\] 
Therefore ${\rm Im}G_{nks}(\epsilon) = -\pi\delta(\epsilon - \xi_{ks})$, which removes 
the integral over $\epsilon$ in (\ref{Pi}). It is convenient to represent ${\cal G}_{nks}$ in
the form 
\begin{equation}
{\cal G}_{nks}(\epsilon) = {\cal G}_{nks}^0(\epsilon) 
 + {\cal G}_{nks}^0(\epsilon) {\cal T}_{nks}(\epsilon) {\cal G}_{nks}^0(\epsilon) ,
\label{T_ns}
\end{equation}
where ${\cal T}_{nks}(\epsilon)$ is a diagonal element of the $\cal T$-matrix. Since 
the interaction in $H(\psi)$ is local, ${\cal T}_{nks}(\epsilon)$ is in fact independent of $k$:
${\cal T}_{nks} = {\cal T}_{ns}$. 
Obviously, for $\epsilon = \xi_{ks} +\omega$ the second term in the r.h.s. 
of (\ref{T_ns}) diverges as $\omega^{-2}$ at small $\omega$. Combination 
of Eq.~(\ref{T_ns}) with Eqs.~(\ref{kubo1}) and (\ref{Pi}), and replacement 
of the sum over $k$ in Eq.~(\ref{kubo1}) by an integral then yield 
Eq.~(\ref{RESPONSE}). 

\section{The effective exchange Hamiltonian for the general form of 
intradot interactions}
\label{E&N}
Two electrons occupying two single particle energy levels can form three 
linearly independent singlet states: 
\begin{eqnarray*}
&&
|0,0\rangle_{-1} 
= d_{-1\uparrow}^\dagger d_{-1\downarrow}^\dagger |0\rangle, \\
&&
|0,0\rangle_{+1} 
= d_{+1\uparrow}^\dagger d_{+1\downarrow}^\dagger |0\rangle, \\
&&
|0,0\rangle_{0}
= \frac{1}{\sqrt{2}}\left( d_{+1\uparrow}^\dagger d_{-1\downarrow}^\dagger 
- d_{+1\downarrow}^\dagger d_{-1\uparrow}^\dagger \right) | 0\rangle .  
\end{eqnarray*}
These states are the eigenstates of the Hamiltonian (\ref{Hdot}) with
$|0,0\rangle_{-1}$ being the singlet state of the 
lowest energy. However, (\ref{Hdot}) is oversimplified by the assumption 
of the validity of the Random Matrix Theory \cite{exchange}. In a more 
general model electron-electron interactions yield non-zero matrix elements
\[
_n\langle 0,0| H_{dot}|0,0\rangle_{n'}\neq 0
\]
for $n\neq n'$. As the result, the lowest energy singlet $|0,0\rangle$ 
is no longer $|0,0\rangle_{-1}$, but a certain mixture of all 3 states 
$|0,0\rangle_n$: 
\[
|0,0\rangle = \sum_{n=0,\pm 1} C_n |0,0\rangle_n .
\quad
\sum_n C_n^2=1 
\]
This generalization of the dot's Hamiltonian yields additional coupling constants 
in the dot-lead exchange interaction. It turns out that such an extended model is 
identical to Eq.~(\ref{Hn}) as far as the low-temperature properties of the system 
are concerned.
We demonstrate it on a specific example \cite{EN,EN1}:
\[
C_0 = 0, 
\quad 
C_n = \frac{n}{\sqrt 2} (-\cos\alpha +n\sin\alpha), 
\]
The case considered in \cite{EN,EN1} corresponds to $\alpha =0$ 
(or $C_{-1}=-C_1 = 1/\sqrt 2$). The choice of the singlet state in \cite{PG} and 
in Eq.~(\ref{Basis}) above corresponds to $\alpha =\pi/4$ (or $C_{-1}=1,\,C_1=0$).

Of course, the transitions between the state $|0,0\rangle$  and the triplet states 
$|1,S^z\rangle$ can be described again in terms of 2 fictitious spin-1/2 operators 
${\bf S}_{1,2}$. But instead of (\ref{Snn})-(\ref{Cnn}) we have now 
\begin{eqnarray*}
{\cal P}\sum_{ss'}d_{ns}^{\dagger}\frac{{\bbox \sigma }_{ss'}}{2}&& d_{n's'}{\cal P} 
= \frac{1}{2}\delta_{n,n'} {\bf S}_{+} 
\\
&& + \frac{1}{2}\delta_{n,-n'}({\bf S}_{-}\cos\alpha  + 2i n{\bf T}\sin\alpha) ,
\\
{\cal P}\sum_s d_{ns}^{\dagger }d_{n's}&&{\cal P} 
=  n\delta_{n,n'} 
\left[\left({\bf S}_1 \cdot {\bf S}_2 \right)\sin (2\alpha)  - 1/4 + n \right] 
\end{eqnarray*}
Accordingly, the effective Hamiltonian also modifies: Eq.~(\ref{model}) 
remains unchanged while Eq.~(\ref{Hn}) is replaced by
\begin{eqnarray}
&&H_{n} = J ({\bf s}_{nn}\cdot {\bf S}_{+}) 
+ V' n\rho_{nn}( {\bf S}_{1}\cdot {\bf S}_{2})   
\nonumber \\
&&\quad\quad\quad\quad
+ \frac{I_S}{\sqrt 2}({\bf s}_{-n,n}\cdot {\bf S}_{-})
+ \frac{I_T}{\sqrt 2}2in({\bf s}_{-n,n}\cdot {\bf T}) .   
\label{ENmodel} 
\end{eqnarray}
The coupling constants here are expressed through $J,I,V$ of Eq.~(\ref{bare}) as
\[
I_S = I\sqrt{2}\cos\alpha ,
\quad
I_T = I\sqrt{2}\sin\alpha ,
\quad
V' = V\sin 2\alpha 
\]
[$J$ is the same as in (\ref{bare})].

The scaling equations for this Hamiltonian read
\begin{eqnarray}
&&
dJ/d{\cal L} = \nu \left[J^2 
+ \left( I_S ^2 + I_T^2 \right)/2 \right] 
\nonumber \\
&&
dI_{S} /d{\cal L} 
= 2\nu \left( I_S J + I_T V' \right) 
\nonumber\\
&&
dI_{T} /d{\cal L} 
= 2\nu \left( I_T J + I_S V' \right) 
\label{Scaling_Interactions}\\
&&
dV'/d{\cal L} = 2\nu I_S I_T \nonumber
\end{eqnarray}
(these equations are equivalent to Eqs.~(17)-(20) of\cite{EN}). 
For $\alpha\neq 0$ the initial values of all coupling constants differ from zero. 
System of equations (\ref{Scaling_Interactions}) allows for a solution in which 
all the coupling constants diverge at a certain point ${\cal L}={\cal L}_0$. Similar 
to the conventional Kondo problem, the leading divergency is 
$J,I_S,I_T,V' \propto \left({\cal L}_0 - {\cal L} \right)^{-1}$.
Examination of the subleading terms allows us to conclude that 
the solutions of (\ref{Scaling_Interactions}) and (\ref{Scaling}) coincide near 
this point if one sets $I_S=I_T = I$ and $V'=V$. 
Indeed, it follows from the second and the third equations in (\ref{Scaling_Interactions}) 
that
\begin{equation}
\frac{d}{d \cal L} \ln (I_-/I_+) = - 2\nu V',
\quad
I_{\pm} = \frac12(I_S \pm I_T)
\label{Ipm}
\end{equation}
Using the leading term of the asymptote of $V'$,
\[
1/\nu V' = (2/\zeta) \left({\cal L}_0 - {\cal L} \right),\quad \zeta >0,
\]
and Eq.~(\ref{Ipm}) we find $I_-/I_+ \propto \left({\cal L}_0 - {\cal L} \right)^\zeta$.
Therefore, in the vicinity of the point ${\cal L}_0$ the difference between 
$I_S$ and $I_T$ can be neglected, and the resulting equations for $J,I_+,V'$ 
coincide with Eqs.~(\ref{Scaling}) for $J,I,V$. 

So far we found a solution near some singularity point ${\cal L}={\cal L}_0$. 
We should check now that such point is reached indeed by solutions satisfying the
proper initial conditions at ${\cal L}=0$. We were only able to perform this check by
solving numerically the system (\ref{Scaling_Interactions}). Numerical solution for a 
set of models defined by $\alpha$ in the interval $0.01 \pi\leq\alpha \leq \pi/4$ 
indeed verified the proper divergency of the coupling constants.
We have also found that the parameter $c$ in Eq.~(\ref{T0}) approaches $1/2$ 
at small $\alpha$. In other words, in the limit $\alpha \rightarrow +0$ the 
difference between $T_0$ and $T_K^{\rm odd}$ disappears.

The convergence of the system of equations (\ref{Scaling_Interactions})  to 
Eqs.~(\ref{Scaling}) means that the results of the main text are not sensitive to the
details of intradot interactions. In particular, Eq.(\ref{TKondo_triplet}) with $\lambda$ 
given by (\ref{lambda}) holds.

We now consider the {\it special case} $\alpha = 0$, in which the exponent $\lambda$ 
turns out to be different. We consider a slight generalization of the model, allowing 
for the dependence of the tunneling amplitudes on the orbital index: 
$t_{\alpha}$ in  Eq.~(\ref{tunneling}) is replaced by $t_{\alpha n}$; 
see also Appendix~\ref{STAIRCASE}). [Note that for $\alpha \neq 0$ this extension 
does not affect the value of $\lambda$ in Eq.~(\ref{lambda})]. 
The exchange Hamiltonian for such model,
\begin{equation}
H_{n} = J_n ({\bf s}_{nn}\cdot {\bf S}_{+}) 
+ I({\bf s}_{-n,n}\cdot {\bf S}_{-}),
\label{EN}
\end{equation}
acquires an $n$-dependent exchange constant. 
The scaling equations in this case,
\begin{equation}
\frac{d J_n }{d\cal L} = \nu (J_n ^2 +I^2), 
\quad 
\frac{d I}{d\cal L} = \nu I\sum_n J_n , 
\label{ENscaling}
\end{equation}
have first integral 
\[
(J_{+1} - J_{-1})/{2I} = \tan \varphi , \quad |\varphi|<\pi/2 .
\]
The value of $\varphi$ is determined by the bare value of the exchange constants.  
The first integral allows one to solve the scaling equations 
(\ref{ENscaling}) exactly \cite{EN,EN1}. Near the singularity point ${\cal L}_0$ 
the solutions acquire asymptotic form
\begin{equation}
\nu J_n = \frac{1+ n \sin\varphi}{2({\cal L}_0 - {\cal L})},
\quad
{\cal L}_0 = \ln(\delta/T_0) .
\label{ENsolution}
\end{equation}
At the triplet side of the transition at energies below $|K|\gg T_0,E_Z$ the system 
is described by the Hamiltonian (\ref{Htriplet}) with $n$-dependent exchange 
amplitudes $J_n$. In this situation one can define two separate scales, 
$T_n$, corresponding to the two exchange constants,
\[
T_n = |K| \exp\left[-1/\nu J_n ({\cal L}^*)\right],
\quad 
{\cal L}^* = \ln(\delta/|K|) .
\]
At these scales the two exchange constants, $\nu J_n$, become of the order of 1.
Substitution here of $J_n({\cal L})$ from Eq.~(\ref{ENsolution}) then yields
\[
\frac{T_n}{T_0} = \left(\frac{T_0}{|K|}\right)^{\lambda_n},
\quad
\lambda_n(\varphi) = \frac{1- n \sin\varphi}{1 + n \sin\varphi}. 
\]
The larger of the two scales corresponds to the smallest of the two 
exponents $\lambda_n$ \cite{EN},\cite{EN1}: 
\[
\min \lambda_n = \frac{1- \sin|\varphi|}{1 +  \sin|\varphi|} < 1. 
\]

The Hamiltonian (\ref{EN}) formally coincides with that for the two impurity 
Kondo model \cite{ALJ}, see Eq.~(\ref{2IKM}) above. If $J_n$ do not depend of $n$, 
this model exhibits an exotic non-Fermi liquid behavior \cite{ALJ}. Note that the 
non-Fermi-liquid state is extremely fragile and is also destroyed by potential scattering, 
which we have neglected in our derivation. The requirements $\alpha =0$ and $\varphi =0$ 
set strict conditions on both the intradot interaction constants and on the values of the 
tunneling amplitudes. Finding these conditions is beyond the scope of the present paper. 

\section{The effective exchange Hamiltonian for the case of orbital mixing}
\label{MIXING}

In order to see how our results are modified in the case when the orbital mixing 
is allowed in the course of tunneling, we choose the tunneling amplitudes in 
Eq.~(\ref{HT}) in the form
\begin{equation}
t_{\alpha nm}=t_{\alpha }\delta _{n,m}+t_{\alpha }'\delta _{n,-m},
\quad \alpha = R,L
\label{tunneling3}
\end{equation}
The rotation in the orbital space,
\begin{eqnarray*}
c_{\alpha nks} & = & \frac{1}{\sqrt{2}}
\left( \widetilde{c}_{\alpha nks}-n\widetilde{c}_{\alpha -nks}\right) ,
\nonumber \\
d_{ns} & = & \frac{1}{\sqrt{2}}
\left(\widetilde{d}_{ns}-n\widetilde{d}_{-ns}\right) 
\end{eqnarray*}
diagonalizes $H_{T}$:
\[
H_{T}=\sum_{\alpha n}v_{\alpha n}
\widetilde{c}_{\alpha nks}^{\dagger }\widetilde{d}_{ns}+{\rm H.c.},
\quad
v_{\alpha n}=t_{\alpha }+nt_{\alpha}'
\]
and modifies Eqs.~(\ref{Snn}),(\ref{Cnn}). 
Next, we perform a rotation in the $R-L$ space, similar to (\ref{rotation})
\[
\left(
\begin{array}{c}
\psi _{nks} \\
\phi _{nks}
\end{array}
\right) 
= \frac{1}{\sqrt {v_{Ln}^{~2}+v_{Rn}^{~2}}}
\left(
\begin{array}{cc}
v_{Rn} & v_{Ln} \\
-v_{Ln} & v_{Rn}
\end{array}
\right) \left(
\begin{array}{c}
\widetilde{c}_{Rnks} \\
\widetilde{c}_{Lnks}
\end{array}
\right) ,
\]
which yields
\[
H_{T}=\sum_{nks}v_{n}
\psi _{nks}^{\dagger }\widetilde{d}_{ns}+{\rm H.c.},
\quad
v_{n}^{2} = v_{Ln}^{~2} + v_{Rn}^{~2},
\]
We now perform a Schrieffer-Wolff transformation, which yields Eq.~(\ref{model}) 
with $H_n$ given by
\begin{eqnarray}
H_{n} & = & J_n\left({\bf s}_{n,n}\cdot {\bf S}_{+}\right)
+ I_n \frac{n}{\sqrt{2}}\left({\bf s}_{n,n}\cdot {\bf S}_{-}\right)
\nonumber
\\
&& \quad + I\sqrt{2}in\left( {\bf s}_{-n,n}\cdot {\bf T}\right)
- V\rho _{-n,n}\left( {\bf S}_1\cdot {\bf S}_2 \right) .
\label{Hn1}
\end{eqnarray}
The scaling equations for this model read
\begin{eqnarray}
dJ_n/d{\cal L} & = &
\nu \left[ J_n^2 + \left( I_n^2+I^2 \right) /2 \right]
\nonumber \\
dI_n/d{\cal L} & = & 2\nu \left( IV+J_n I_n \right)
\nonumber \\
dI/d{\cal L} & = & \nu \sum_n \left( IJ_n  +VI_n\right)  
\label{RGmixing}\\
dV/d{\cal L} & = & \nu \sum_n II_n 
\nonumber
\end{eqnarray}
In a generic case $t_{\alpha} \neq t'_{\alpha}$ the initial values of 
all the coupling constants in (\ref{RGmixing}) are finite:
\[
J_n=I_n=2v_n^2/E_C,
\;
I=2V=2v_{+1}v_{-1}/E_C.
\]
For example, for 
\begin{equation}
t_\alpha'/t_\alpha = \beta
\label{beta}
\end{equation}
one obtains
\[
J_n = I_n = (1+n\beta)^2 J_0, \quad I = 2V = (1- \beta ^2) J_0, 
\]
where $J_0 = 2(t_L^2 + t_R^2)/E_C$.

System of equations (\ref{RGmixing}) allows for a solution in which 
all the coupling constants diverge at a certain point ${\cal L}={\cal L}_0$, 
with the leading divergency being of the form 
$J_n,I_n,I,V \propto \left({\cal L}_0 - {\cal L} \right)^{-1}$. 
The analysis of the subleading terms similar to that in Appendix~\ref{E&N} 
leads to a conclusion that in the vicinity of the point ${\cal L}_0$ the difference 
between $J_{+1}$ and $J_{-1}$, and the difference between $I_n$ and $I$ 
can be neglected. The resulting equations are identical to Eqs.~(\ref{Scaling}). 
Accordingly, near the point ${\cal L}_0$ the solutions of Eqs.~(\ref{RGmixing}) 
coincide with that of Eqs.~(\ref{ScalingSolution}) if $J_n=J$ and $I_n=I$. 
Note that setting $J_n=J$ and $I_n=I$ in Eq.~(\ref{Hn1}) and performing 
a unitary transformation 
\[
\psi _{nks}=\frac{1}{\sqrt{2}}\left( a_{nks}+na_{-nks}\right) 
\]
brings the exchange Hamiltonian (\ref{Hn1}) to the form (\ref{Hn}). 
 
We solved Eqs.~(\ref{RGmixing}) numerically for $\beta$ [see Eq.~(\ref{beta})] in the 
interval $0\leq\beta\leq 0.9$ (note that for the model defined by Eqs.~(\ref{tunneling3}) 
and (\ref{beta}) all physical observables possess a symmetry with respect to
the change $\beta\rightarrow 1/\beta$). These calculations confirmed that the proper divergency 
of the coupling constants indeed occurs.  

It turns out that the relation between $T_K^{\rm triplet}$ and $T_K^{\rm odd}$ 
(see Section~\ref{SCALING}) depends on the degree of orbital mixing: the stronger 
the orbitals are mixed the less the difference between $T_K^{\rm triplet}$ and 
$T_K^{\rm odd}$ is. For example, for the model (\ref{tunneling3}),(\ref{beta}) 
$T_K^{\rm triplet}$ can be associated with the larger of the two energy scales 
$T_n = \delta \exp (-1/\nu J_n)$, while $T_K^{\rm odd} = \delta \exp (-1/\nu J_{\rm odd})$ 
with $J_{\rm odd} = 2(1 + \beta^2)J_0$. This yields
\[
\frac{\ln\left(\delta/T_K^{\rm triplet}\right)}{\ln\left(\delta/T_K^{\rm odd}\right)} 
= 2\frac{1+\beta^2}{(1+\beta)^2}.
\]
The expression in the r.h.s. here is invariant with respect to the change 
$\beta\leftrightarrow 1/\beta$ and its value varies between $2$ for $\beta =0$ 
(or $\beta =\infty$) and $1$ for $\beta = 1$. Therefore, the temperature scales 
$T_K^{\rm triplet}$ and $T_K^{\rm odd}$ coincide in the strong mixing limit 
$\beta \rightarrow 1$. The experiments with vertical quantum dots \cite{sasaki} 
correspond to $T_K^{\rm triplet} \ll  T_K^{\rm odd}$, which can be explained 
only if the mixing is weak ($\beta \ll 1$ or $1/\beta \ll 1$). 

Note that the relation between $T_0$ and $T_K^{\rm odd}$ depends on $\beta$ 
much weaker. For instance, for $\beta=1$ the ratio 
\[
\frac{\ln\left(\delta/T_0\right)}{\ln\left(\delta/T_K^{\rm odd}\right)} 
= \frac{1}{1+1/\sqrt 2}\approx 0.6,
\] 
while for $\beta=0$ this ratio equals approximately $0.7$, see Eqs.~(\ref{T0}) and 
(\ref{TK_odd}). Accordingly, for $\delta \gg T_0$ the inequality 
$T_0\gg T_K^{\rm odd}$ holds independently of the strength of the orbital mixing. 

\section{Two--stage Kondo effect}
\label{STAIRCASE}

In this Appendix, we consider an example of the non-universal features that 
the system may exhibit far away from the transition point. Consider the following 
generalization of (\ref{tunneling}):
\begin{equation}
t_{\alpha nn'} = t_{\alpha n}\delta_{nn'}.
\label{tunnelingNEW}
\end{equation}
After a rotation in the left-right space
\[
\left( 
\begin{array}{c}
\psi _{nks} \\ 
\phi _{nks}
\end{array}
\right) = 
\frac{1}{t_n} 
\left( 
\begin{array}{cc}
t_{Rn} & t_{Ln} \\ 
-t_{Ln} & t_{Rn}
\end{array}
\right)
\left( 
\begin{array}{c}
c_{Rnks} \\ 
c_{Lnks}
\end{array}
\right),
\;
t_n^2 = t_{Ln}^2 + t_{Rn}^2
\]
and a Schrieffer-Wolff transformation one obtains Eqs.~(\ref{model})-(\ref{Hn}) 
with $J$ and $V$ in (\ref{Hn}) replaced by $J_n$ and $V_n$, 
\[
J_n = 2V_n = 2t_n^2/E_C, 
\quad 
I= 2 t_{+1}t_{-1}/E_C.
\] 
This is only a small modification as far as the vicinity of the singlet-triplet transition
is concerned. Indeed, the scaling equations (\ref{Scaling}) are replaced by
\begin{eqnarray*}
&&
dJ_n/d{\cal L} = \nu \left( J_n^2 + I^2 \right), \\
&&
dI/d{\cal L} = \nu I\sum_n\left( J_n + V_n \right) , \\
&&
dV_n/d{\cal L} = 2\nu I^2 
\end{eqnarray*}
If these equations are rewritten in terms of new variables
\begin{eqnarray*}
&& J = (J_{+1}+J_{-1})/2, \;  V = (V_{+1}+V_{-1})/2, \\
&& J' = (J_{+1}-J_{-1})/2 , \;  V' = (V_{+1}-V_{-1})/2,
\end{eqnarray*}
one finds that $J,I$, and $V$ behave at large $\cal L$ according to 
Eq.~(\ref{ScalingSolution}), all being proportional to 
$\left({\cal L}_0 - {\cal L}\right)^{-1}$,  while $J'$ diverges much slower, as
$J' \propto \left({\cal L}_0 - {\cal L}\right)^{-2/(\lambda + 1)}$, and $V'$ 
is not renormalized at all.

\begin{figure}[tbp]
\centerline{\epsfxsize=6cm
\epsfbox{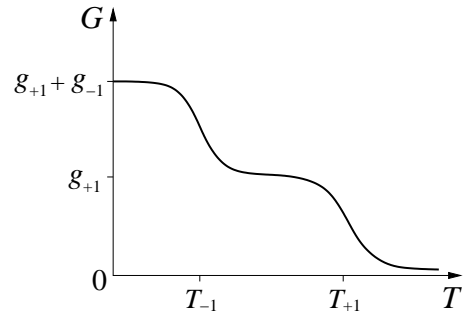}\vspace{1.5mm}}
\caption{Staircase-like temperature dependence of the linear conductance at the 
triplet side of the singlet-triplet transition.}
\label{staircase}
\end{figure}

Interesting properties appear at the triplet side of the transition in the case when the 
exchange constants are very different, say, $J_{+1}\gg J_{-1}$.
 
The exchange term in the Hamiltonian (\ref{Htriplet}) is
replaced by $\sum_n J_n \left({\bf s}_{nn}\cdot {\bf S}\right)$.
One can define two separate Kondo temperatures 
\[
T_n = \delta \exp(-1/\nu J_n), 
\quad T_{+1} \gg T_{-1}  
\]
corresponding to the two exchange constants $J_n$. 
At $T\gg T_{+1}\gg T_{-1}$ the presence of the $n=-1$ channel  can 
be neglected, so that the system in this regime is described by a single-channel $S=1$ 
Kondo model.  Accordingly, the conductance in this regime is given by 
\begin{equation}
G = g_{+1}\frac{\pi^2/2}{[\ln(T/T_{+1})]^2},
\quad T\gg T_{+1}
\label{stair1}
\end{equation}
where $g_n = (2e^2/h) 4t_{Ln}^2t_{Rn}^2 t_{n}^{-4}$. 
At $T\ll T_{+1}$ the scattering in the $n=+1$ channel reaches the unitary limit. 
In this regime one half of the dot's spin is screened \cite{NB}. The remaining spin 
$S=1/2$ still interacts with the electron continuum in the channel $n=-1$. This 
antiferromagnetic exchange coupling eventually results in the second stage of the 
Kondo effect.  The corresponding  asymptote of the conductance is
\begin{equation}
G = g_{+1} + g_{-1} \frac{3\pi^2/16}{[\ln(T/T_{-1})]^2},
\quad T_{-1}\ll T\ll T_{+1}
\label{stair2}
\end{equation}
At $T \ll T_{-1}$ the dot's spin is screened entirely \cite{NB}. 
The zero-temperature conductance $G_0 = g_{+1} + g_{-1}$; the 
leading finite temperature correction $G_0 - G \propto (T/T_{-1})^2$. 

Equations (\ref{stair1}) and (\ref{stair2}) describe a staircase-like temperature 
dependence of the conductance, see Fig.~\ref{staircase}. The thermodynamic 
quantities are expected to exhibit a similar two-stage crossover \cite{AJ}.

\end{multicols}
\end{document}